\documentclass[11pt]{article}

\input{stylesheet.sty}

\usepackage{booktabs} 
\usepackage[ruled,linesnumbered]{algorithm2e} 

\usepackage{algpseudocode}

\usepackage[T1]{fontenc}

\usepackage{cleveref}
\crefname{algocf}{Algorithm}{Algorithms} 
\Crefname{algocf}{Algorithm}{Algorithms} 

\newcommand{\lc}{\left\lceil}
\newcommand{\rc}{\right\rceil}

\SetAlFnt{\small}
\SetAlCapFnt{\small}
\SetAlCapNameFnt{\small}
\SetAlCapHSkip{0pt}
\IncMargin{-\parindent}

\usepackage{tikz}
\usetikzlibrary{positioning}

\usepackage{hyperref} 
\hypersetup{
    colorlinks=true,
    allcolors=teal,
    pdfencoding=auto,
}

\usepackage{cleveref}

\title{On Hierarchies of Fairness Notions in Cake Cutting: From Proportionality to Super Envy-Freeness}

\newcommand{\alloc}{A}

\newcommand{\newnotionlinear}{Complement Linearly Bounded}

\newcommand{\newnotionlinearabbrev}{\textsc{CLB}}

\newcommand{\newnotioninverse}{Complement Harmonically Bounded}

\newcommand{\newnotioninverseabbrev}{\textsc{CHB}}

\newcommand{\newalgosecondstage}{Last Diminisher}

\author{
Arnav Mehra\thanks{Capital One. Work done while at Purdue University. Email: mehraarn000@gmail.com} \and 
Alexandros Psomas\thanks{Purdue University. Email: apsomas@purdue.edu}
}

\date{}

\begin{document}

\begin{titlepage}
 
\maketitle
\thispagestyle{empty}

\begin{abstract}
We consider the classic cake-cutting problem of producing fair allocations for $n$ agents, in the Robertson–Webb query model. In this model, it is known that: (i) proportional allocations can be computed using $O(n \log n)$ queries, and this is optimal for deterministic protocols; (ii) envy-free allocations (a subset of proportional allocations) can be computed using $O\left( n^{n^{n^{n^{n^{n}}}}} \right)$ queries, and the best known lower bound is $\Omega(n^2)$; (iii) perfect allocations (a subset of envy-free allocations) cannot be computed using a bounded (in $n$) number of queries.

In this work, we introduce two hierarchies of new fairness notions: Complement Harmonically Bounded 
 (CHB) and Complement Linearly Bounded (CLB). An allocation is CHB-$k$, if for any subset of agents $S$ of size at most $k$, and every agent $i \in S$, the value of $i$ for the union of all pieces allocated to agents outside of $S$ is at most $\frac{n-|S|}{n-|S|+1}$; for CLB-$k$ allocations, the upper bound becomes $\frac{n-|S|}{n}$. Intuitively, these notions of fairness ask that, for every agent $i$, the collective value that a group of agents has (from the perspective of agent $i$) is limited.
 CHB-$k$ and CLB-$k$ coincide with proportionality for $k=1$. For all $k \leq n$, CHB-$k$ allocations are a superset of envy-free allocations (i.e., easier to find). On the other hand, for $k \in [2, \lceil n/2 \rceil - 1]$, CLB-$k$ allocations are incomparable to envy-free allocations. For $k \geq \lceil n/2 \rceil$, CLB-$k$ allocations are a subset of envy-free allocations (i.e., harder to find), while CLB-$n$ coincides with super envy-freeness: the value of each agent for her piece is at least $1/n$, and her value for the piece allocated to any other agent is at most $1/n$.

 We prove that CHB-$n$ allocations can be computed using $O(n^4)$ queries in the Robertson–Webb model. On the flip side, finding CHB-$2$ (and therefore all CHB-$k$ for $k \geq 2$) allocations requires $\Omega(n^2)$ queries, while CLB-$2$ (and therefore all CLB-$k$ for $k \geq 2$) allocations cannot be computed using a bounded (in $n$) number of queries. 
 Our results show that envy-free allocations occupy a curious middle ground, between a computationally impossible notion of fairness, CLB-$\lceil n/2 \rceil$, and a computationally ``easy'' notion, CHB-$n$.
\end{abstract}

\end{titlepage}

\section{Introduction}

We consider the classic problem of cake cutting, proposed by Hugo Steinhaus~\cite{steinhaus1948problem}, while in hiding during World War II~\cite{wikipedia_steinhaus}. In this problem, there is a heterogeneous resource, the ``cake,'' typically represented by the interval $[0,1]$. There is a set $N$ of $n$ agents with different valuations functions $V_1, \dots, V_n$ over subsets of the cake; these valuations functions are induced by probability measures over $[0,1]$ (and hence, for example, the value of every agent for the entire cake is equal to $1$). The goal in this problem is to allocate to each agent a piece of the cake --- a finite union of disjoint intervals --- in a fair manner, under various notions of fairness. This simple model has served as a cornerstone for fair division, shaping many of the field's foundational questions.

A major focus has been the complexity of computing fair allocations. The complexity of discrete cake-cutting protocols is measured by the number of queries they require in the query model suggested by Robertson and Webb~\cite{robertson1998cake} and later formalized (and named) by Woeginger and Sgall~\cite{woeginger2007complexity}. For example, it is well understood that \emph{proportional allocations}, those where every agent $i \in N$ values her piece at least $1/n$, can be computed using $O(n \log n)$ queries~\cite{even1984note}; and, this is tight: every deterministic protocol requires $\Omega(n \log n)$ queries~\cite{edmonds2011cake}.
Another major notion of fairness is \emph{equitability}, which requires that every agent has the same value for the piece allocated to them, i.e., for all agents $i, j \in N$, $V_i(\alloc_i) = V_j(\alloc_j)$ (where $\alloc_{\ell}$ is the piece allocated to agent $\ell$). Procaccia and Wang~\cite{procaccia2017lower} proved that equitable allocations cannot be computed in a bounded (in $n$) number of queries. This result, in turn, rules out bounded protocols for another major notion of fairness, \emph{perfection}: an allocation $\alloc$ is perfect if for all agents $i,j \in N$, $V_i(\alloc_j) = 1/n$.\footnote{Which, perhaps surprisingly, always exist~\cite{liapounoff1940fonctions,alon1987splitting}!} 

Arguably the most important notion of fairness --- the holy grail of fair division --- is \emph{envy-freeness}; an allocation $\alloc$ is envy-free if every agent prefers her piece to the piece allocated to any other agent, i.e., for all agents $i,j \in N$, $V_i(\alloc_i) \geq V_i(\alloc_j)$. It is easy to see that every perfect allocation is envy-free, and every envy-free and complete ($\cup_i \alloc_i = [0,1]$) allocation is proportional. Unlike proportionality, equitability, and perfection, the complexity of computing (complete) envy-free allocations has been rather elusive. The existence of \emph{bounded} protocols remained open for two decades until Aziz and Mackenzie~\cite{aziz2016discrete} presented a protocol that requires at most $O(n^{n^{n^{n^{n^{n}}}}})$ queries. The best known lower bound is $\Omega(n^2)$~\cite{procaccia2009thou}, leaving an astronomical gap in our understanding.

In this paper, we zoom in on the landscape of fairness between proportionality and perfection, focusing on the query complexity of different notions in this less understood intermediate space. 




\subsection{Our contributions}

We introduce two new hierarchies of fairness: \newnotioninverse ~(\newnotioninverseabbrev) and \newnotionlinear ~(\newnotionlinearabbrev). An allocation $\alloc$ satisfies $\newnotioninverseabbrev\text{-}k$ if, for every subset of agents $S$ such that $|S| \leq k$, and for every agent $i \in S$, the value of $i$ for the union of pieces allocated to agents not in $S$ is at most $\frac{n - |S|}{n-|S|+1}$, i.e., $\sum_{j \notin S} V_i(\alloc_j) \leq \frac{n - |S|}{n-|S|+1}$. For $\newnotionlinearabbrev\text{-}k$, the upper bound is $\frac{n - |S|}{n}$ instead. The existence of $\newnotioninverseabbrev\text{-}k$ allocations is implied by the existence of envy-free allocations. Consider an envy-free allocation $\alloc$; we have that, for all agents $i, j \in N$, $V_i(\alloc_j) \leq V_i(\alloc_i)$. Adding up for all $j \notin S$ we have $\sum_{j \notin S} V_i(\alloc_j) \leq (n- |S|) V_i (\alloc_i) \leq (n- |S|)( 1 - \sum_{j \notin S} V_i(\alloc_j))$; re-arranging gives the desired inequality. The existence of $\newnotionlinearabbrev\text{-}k$ allocations is similarly implied by the existence of perfect allocations.

Our new fairness notions are interesting precisely because they occupy a natural but largely unexplored middle ground between proportionality and envy-freeness. $\newnotioninverseabbrev\text{-}k$ allocations limit how much value groups of agents collectively derive from other agents' shares; thus, they may be particularly suited for contexts where collective perceptions of fairness are important. $\newnotionlinearabbrev\text{-}k$ allocations impose stronger conditions, which makes them appealing when perfection is desired (where the computational cost is prohibitive/unbounded). By providing a spectrum of fairness guarantees, these notions offer a more nuanced toolkit, enabling practitioners to better navigate the trade-off between computational complexity and fairness requirements.

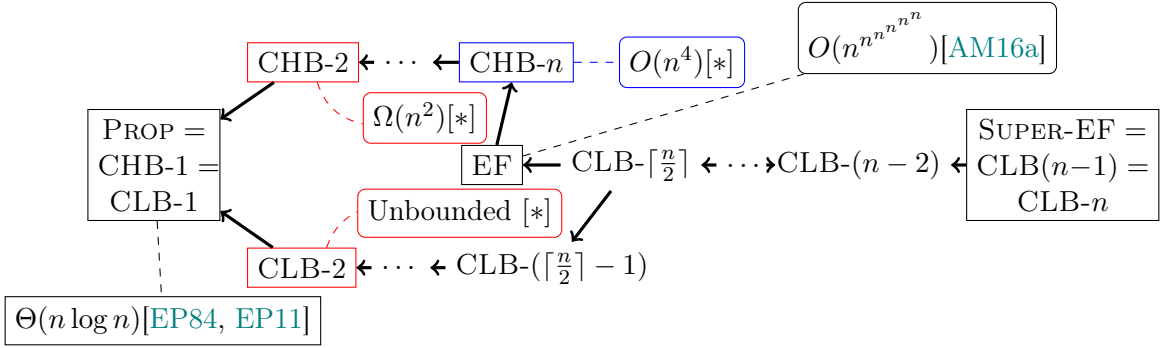
\begin{figure}[t]
\centering
\begin{tikzpicture}[node distance=1.5cm and 1.5cm, scale=0.8]  
  
  \node (prop) [draw, rectangle, text width=1.5cm, align=center] {$\textsc{Prop} = \newnotioninverseabbrev\text{-}1 = \textsc{\newnotionlinearabbrev-}1$};

\node (prop_bounds) [draw, rectangle, below left=1.0cm and 1.1cm of prop, anchor=north west] {$\Theta(n \log n)$\cite{even1984note,edmonds2011cake}};

  \node (chb2) [draw=red, rectangle,above right=0.5cm of prop] {\newnotioninverseabbrev\text{-}2};
  \node (chbdots) [right=0.2cm of chb2] {$\dots$};
  \node (chbn) [draw=blue, rectangle,right of=chbdots] {\newnotioninverseabbrev\text{-}$n$};

  \node (chb2_lower) [draw=red, rectangle, rounded corners=1mm, below right=0.13cm and 0.05cm of chb2] {$\Omega(n^2) [\ast]$};

  \node (chbn_upper) [draw=blue, rectangle, rounded corners=1mm, right=0.6cm of chbn] {$O(n^4) [\ast]$};

  \node (ef_upper) [draw, rectangle, rounded corners=1mm,above right=-0.5cm and 0.7cm of chbn_upper] {$O(n^{n^{n^{n^{n^{n}}}}})$\cite{aziz2016discrete}};

  \node (clb2) [draw=red, rectangle,below right=0.5cm of prop] {\textsc{\newnotionlinearabbrev-}2};
  \node (clbdots) [right=0.2cm of clb2] {$\dots$};
  \node (clbceil) [right=0.2 of clbdots] {\textsc{\newnotionlinearabbrev-}$(\lceil\frac{n}{2}\rceil-1)$};

  \node (clb2_lower) [draw=red, rectangle, rounded corners=1mm, above right=0.13 and 0.01cm of clb2] {Unbounded $[\ast]$};

  \node (ef) [draw, rectangle,right=3.2cm of prop] {\textsc{EF}};
  \node (clbceilplus) [right=0.5cm of ef] {\textsc{\newnotionlinearabbrev-}$\lceil\frac{n}{2}\rceil$};
  \node (clbdots2) [right=0.2cm of clbceilplus] {$\dots$};
  \node (clbn) [right of=clbdots2] {\textsc{\newnotionlinearabbrev-}$(n-2)$};
  \node (superef) [draw, rectangle, text width=2.3cm, align=center] [right=0.2cm of clbn] {$\textsc{Super-EF}= \textsc{\newnotionlinearabbrev}(n-1) = \textsc{\newnotionlinearabbrev-}n$};
  
  \draw[<-,very thick] (prop) -- (chb2);
  \draw[<-,very thick] (chb2) -- (chbdots);
  \draw[<-,very thick] (chbdots) -- (chbn);
  \draw[<-,very thick] (chbn) -- (ef);
  
  \draw[<-,very thick] (prop) -- (clb2);
  \draw[<-,very thick] (clb2) -- (clbdots);
  \draw[<-,very thick] (clbdots) -- (clbceil);
  \draw[<-,very thick] (clbceil) -- (clbceilplus);
  
  \draw[<-,very thick] (ef) -- (clbceilplus);
  \draw[<-,very thick] (clbceilplus) -- (clbdots2);
  \draw[<-,very thick] (clbdots2) -- (clbn);
  \draw[<-,very thick] (clbn) -- (superef);

  \draw[blue, -, dashed] (chbn) to (chbn_upper);
  \draw[-, dashed] (ef) to (ef_upper);
  \draw[-, dashed] (prop) to (prop_bounds);
  \draw[red, -, dashed, bend right=30] (chb2) to (chb2_lower);
  \draw[red, -, dashed, bend left=20] (clb2) to (clb2_lower);

\end{tikzpicture}
\caption{Relation between our new notions of fairness, Proportionality (\textsc{Prop}), Envy-Freeness (\textsc{EF}), and Super Envy-Freeness (\textsc{Super-EF}). A solid arrow from $Y$ to $X$ represents that notion $Y$ implies notion $X$, i.e. $X \leftarrow Y$ means $X \supsetneq Y$. $X \leftarrow Y$ implies that finding an allocation with property $Y$ is a harder task.  Our results, highlighted in colored rectangles, are marked with $[\ast]$.}
\label{fig:fairness_hierarchy}
\end{figure}

In~\Cref{sec:relations} we start by showing how our new notions relate to the existing notions of fairness; we focus on complete allocations (as, otherwise, envy-freeness is the easiest notion, since it is satisfied by an empty allocation). We show that the entire $\newnotioninverseabbrev\text{-}k$ hierarchy lies between proportionality and envy-freeness, with $\newnotioninverseabbrev\text{-}1$ being proportionality, but $\newnotioninverseabbrev\text{-}n$ being a strict superset of envy-freeness. On the other hand, the entire $\newnotionlinearabbrev\text{-}k$ hierarchy lies between proportionality and perfection, with $\newnotionlinearabbrev\text{-}1$ also being proportionality, while $\newnotionlinearabbrev\text{-}n$ is equivalent to \emph{super envy-freeness}. An allocation $\alloc$ is super envy-free if, for all agents $i,j \in N$, $V_i(\alloc_i) \geq 1/n \geq V_i(\alloc_j)$. Regarding $\newnotionlinearabbrev\text{-}k$ and envy-freeness, for $2 \leq k \leq \lceil \frac{n}{2} \rceil - 1$, $\newnotionlinearabbrev\text{-}k$ is incomparable to envy-freeness; for $k \geq \lceil \frac{n}{2} \rceil$, $\newnotionlinearabbrev\text{-}k$ implies envy-freeness.
See Figure~\ref{fig:fairness_hierarchy}.

We proceed to study the complexity of computing our new notions in the Robertson-Webb model. In~\Cref{sec: algorithms} we prove our main upper bound,~\Cref{thm: main upper bound}: there exists a protocol that computes complete, $\newnotioninverseabbrev\text{-}n$ allocations, using at most $O(n^4)$ queries in the Robertson-Webb model. The first key observation is that, a complete, proportional and ``near-perfect'' allocation $\alloc = (\alloc_1, \dots, \alloc_n)$ satisfies all the constraints required by $\newnotioninverseabbrev\text{-}n$; concretely, completeness, along with $V_i(\alloc_i) \geq \frac{1}{n}$ and $V_i(\alloc_j) \geq \frac{1}{2n}$ for all $j$, implies $\newnotioninverseabbrev\text{-}n$.
Therefore, it is natural to start with an $\epsilon$-perfect allocation --- which can be computed using only $O(n^3/\epsilon)$ queries~\cite{branzei2015dictatorship} --- and try to adjust it to satisfy the additional proportionality property.
An obstacle in this approach is that, for any $\epsilon > 0$, it might be the case that the $\epsilon$-perfect allocation is such that $V_i(\alloc_j) = \frac{1}{n} + \epsilon$, for all $j \neq i$, making it impossible to satisfy proportionality for $i$, unless we trim other agents' pieces. This obstacle can be bypassed with the following trick: introduce $\ell$ phantom agents, with arbitrary valuation functions, and find an $\epsilon$-perfect partition for the instance with $n + \ell$ agents. Then, the ``real'' agents get almost equal allocations, no one gets a piece whose value is larger than $\frac{1}{n}$, and the allocations of the phantom agents can be combined into a new piece, the residue, to be allocated among the real agents, using, e.g., a clever recursion.
However, intuitively, such a step forces us to find an allocation of the residue that is weighted proportional, where agents have different weights (or, equivalently, find an allocation such that agents' utilities meet specific utility lower bounds). Unfortunately, this leads us to a dead-end: there is no bounded protocol for finding a proportional division with unequal shares~\cite{cseh2020complexity}. The final piece of this puzzle is that pieces returned by the $\epsilon$-perfect call do not need to be matched to specific ``real'' agents, but should be available for anyone to claim.

In more detail, our algorithm,~\Cref{algo:inversen}, starts by finding an $\epsilon$-perfect allocation $B$ for $4n/3$ agents, where the $n/3$ extra agents, agents $n+1, \dots, 4n/3$, have arbitrary valuation functions. Let $B_1, \dots, B_{n}$ be the first $n$ pieces of this allocations; pieces $B_{n+1}, \dots, B_{4n/3}$ are combined into the residue $R$. Importantly, ``real'' agents are \emph{not} allocated any of the $B_j$ pieces (yet).
\Cref{algo:inversen} then proceeds \`a la Dubins-Spanier~\cite{dubins1961cut} (or Last Diminisher~\cite{steinhaus1948problem}). Let $S$ be the subset of the $B_j$ pieces still available.
The algorithm asks every agent $i$ to make a mark on $R$, such that her value to the left of the mark is equal to $1/n - \max_{B \in S} V_i(B)$. The agent $i^*$ with the left-most mark (breaking ties arbitrarily) gets the two corresponding pieces, one from cutting the residue and her favorite piece from $S$, and exits the process. The process is repeated until a single agent $j$ is left, who is allocated the entire remaining residue, and the unique piece left in $S$. We prove that~\Cref{algo:inversen} computes a $\newnotioninverseabbrev\text{-}n$ allocation using $O(n^4)$ queries. Importantly, our query complexity bound relies on the fact that the algorithm of Br{\^a}nzei and Miltersen~\cite{branzei2015dictatorship} for finding $\epsilon$-perfect allocation produces at most $O(n^2)$ intervals.

In~\Cref{sec: lower bounds} we prove our main lower bounds. First, we prove that computing a $\newnotioninverseabbrev\text{-}2$ allocation requires $\Omega(n^2)$ queries (\Cref{theorem:CHB2 lower bound}), while computing a $\newnotionlinearabbrev\text{-}2$ allocation is impossible using a bounded protocol (\Cref{theorem:LHkinfinite}). To prove our first lower bound, \Cref{theorem:CHB2 lower bound}, we first formalize the amount of knowledge an algorithm can have about an agent $i$ after $t$ queries. We use the notion of \emph{active intervals}. Intuitively, if an interval $I$ is active for agent $i$ at step $t$, denoted by $I \in \Pi_i^t$, then $V_i(I)$ is known to the algorithm, but for all $I' \subsetneq I$, an adversary can pick $V_i(I')$ to be anything from $0$ to $V_i(I)$. That is, all these choices for $V_i(I')$ are consistent with the query responses up until time $t$. This definition was also at the core of the $\Omega(n^2)$ lower bound for envy-freeness, by Procaccia~\cite{procaccia2009thou}. Here, we argue that, if the adversary responds to queries as if the valuation functions are uniform for the entirety of an algorithm's execution, the allocation $\alloc = (\alloc_1, \dots, \alloc_n)$ that the algorithm outputs must be such that $\alloc_i \in \Pi_i^T$ and $V_i(\alloc_i) = \frac{1}{n}$, for all $i \in N$, where $T$ is the number of queries the algorithm made before terminating. For $\newnotionlinearabbrev\text{-}2$ to be satisfied, $V_i(\alloc_i) = \frac{1}{n}$ implies a condition on $V_i(\alloc_j)$, that similarly needs to be ``checked'' (i.e., $I \in \Pi_i^T$, for some $ I \subseteq \alloc_j$) for all pairs of agents $i,j \in N$; the $\Omega(n^2)$ lower bound follows. Our second lower bound,~\Cref{theorem:LHkinfinite}, reduces the problem of finding an $\newnotionlinearabbrev\text{-}2$ allocation to a known problem with unbounded query complexity: the problem of finding an exact division. In the exact division problem, we are given target values $w_1, \dots, w_z$, and are asked to split the cake into $z$ pieces, so that every agent $i$ has value $w_{\ell}$ for piece $\ell$, for $\ell = 1, \dots, z$. Exact division is known to be impossible to solve with a bounded protocol, even for the case of two valuation functions and $z=2$ pieces with equal weights~\cite{robertson1998cake}. 
~\Cref{theorem:LHkinfinite} immediately implies that the problem of finding a super-envy free allocation has unbounded query complexity in the Robertson-Webb model, which, to the best of our knowledge, was an open problem.

Finally, in~\Cref{sec: approximations}, in light of the strong lower bounds for $\newnotionlinearabbrev\text{-}2$, we consider approximations. The minimally weaker notion that we could hope to achieve is the following: (i) if $|S| = 1$, then $V_i(\alloc_{\bar{S}}) \leq \frac{n-1}{n}$, for all $i \in S$ (i.e., the allocation is proportional), (2) if $|S|=2$, then $V_i(\alloc_{\bar{S}}) \leq \frac{n-2}{n}(1 + \delta)$, for all $i \in S$. We show that this notion is possible to achieve with a bounded protocol. More generally, we define $\delta\text{-}\newnotionlinearabbrev\text{-}n$ to be the set of complete and proportional allocations, such that $k \geq |S| \geq 2$, then $V_i(\alloc_{\bar{S}}) \leq \frac{n-|S|}{n}(1 + \delta)$; we prove that the query complexity of computing a $\delta$-\textsc{\newnotionlinearabbrev}-$n$  allocation is $O\left(\frac{n^6}{\epsilon} \, \frac{\ln(n/\epsilon)}{\ln(n)} \right)$. To prove this result, we give an algorithm for finding complete, proportional, and $\epsilon$-perfect allocations using $O\left(\frac{n^5}{\epsilon} \, \frac{\ln(1/\epsilon)}{\ln(n)} \right)$ queries, which might be of independent interest. The high-level blueprint of this algorithm,~\Cref{algo:almostlinearn}, is similar to~\Cref{algo:inversen}: in phase one split the cake into approximately equal pieces and residue, and in phase two run a ``cut-and-match Last Diminisher.''~\Cref{algo:almostlinearn} requires much stricter conditions from the pieces and residue at the end of phase one. As opposed to~\Cref{algo:inversen}, phase one in~\Cref{algo:almostlinearn} is implemented via a recursion: at each iteration the residue is split into an $(n+1)$-piece $\epsilon'$-perfect allocation. Out of these $n+1$ pieces, one of them serves as the residue in the next iteration. The first $n$ pieces are carefully matched to previously computed pieces. This process is done $d$ times in total, where $d$ and $\epsilon'$ need to be carefully chosen so that the value of every agent for every ``combined'' piece (across iterations) is within the required bounds.

Our results provide significant insight into the spectrum of fairness notions between proportionality and super envy-freeness; see Figure~\ref{fig:fairness_hierarchy}. 
We observe that the two ways of strengthening proportionality, $\newnotioninverseabbrev\text{-}2$ and $\newnotionlinearabbrev\text{-}2$, lead to strikingly different lower bounds on the corresponding query complexity. Envy-freeness occupies a curious middle ground. The minimally weaker fairness notion, $\newnotioninverseabbrev\text{-}n$, can be solved in $O(n^4)$ queries, while the minimally stronger fairness notion, $\newnotionlinearabbrev\text{-}\lceil \frac{n}{2} \rceil$, has unbounded query complexity. A concrete take-home message of our work regarding the query complexity of finding envy-free allocations is that, if a super polynomial lower bound exists, then there must exist a subproblem strictly harder than finding $\newnotioninverseabbrev\text{-}n$ allocations which requires a super polynomial number of queries to solve.

\subsection{Related work}

As already discussed, proportional allocations can be computed using $O(n \log n)$ queries~\cite{even1984note}, and this is tight for deterministic protocols~\cite{edmonds2011cake}. Randomized algorithms can bypass this lower bound~\cite{4031397}. 
Surprisingly, finding a (weighted) proportional allocation when agents have unequal shares requires an unbounded number of queries\cite{cseh2020complexity}.

Regarding envy-freeness, the cut-and-choose method gives an envy-free allocation for $n=2$ agents, and the Selfridge–Conway procedure gives an envy-free allocation for $n=3$ agents. The problem of finding an envy-free allocation for four agents, using a bounded number of queries, was resolved by~\cite{aziz2016discretefour}, and later improved by~\cite{amanatidis2018improved}. For $n$ agents, the best known protocol requires $O(n^{n^{n^{n^{n^{n}}}}})$ queries, while the currently best known lower bound is $\Omega(n^2)$~\cite{procaccia2009thou}. 
Equitable allocations, and therefore perfect allocations, cannot be found in a bounded number of queries~\cite{procaccia2017lower}.

Approximate solutions are typically easier to find.
Br{\^a}nzei and Miltersen~\cite{branzei2015dictatorship} prove that 
finding an $\epsilon$-perfect allocation requires at most $O(n^3/\epsilon)$ queries, and Br{\^a}nzei and Nisan~\cite{branzei2022query} prove that finding an $\epsilon$-envy free and connected allocation requires at most $O(n / \epsilon)$ queries (and at least $\Omega(\log (1/\epsilon))$ queries.
Cechl{\'a}rov{\'a} and Pill{\'a}rov{\'a}~\cite{cechlarova2012computability} prove that finding an $\epsilon$-equitable and proportional allocation requires at most $O(n( \log n + \log (1 / \epsilon))$ queries.

Segal-Halevi and Suksompong~\cite{segal2020cut,segal2023cutting} 
and Segal-Halevi and Nitzan~\cite{segal2019fair} study cake cutting among groups. Specifically, Segal-Halevi and Suksompong~\cite{segal2020cut,segal2023cutting} study contiguous and envy-free cake cutting among groups (where agents within a group also get a contiguous piece). Segal-Halevi and Nitzan~\cite{segal2019fair} study a problem where a cake must be divided and allocated to (pre-determined) groups of agents, and study the existence and query complexity of various fairness notions in this model. For example, in their unanimous FS allocations the $n$ agents are placed in $k$ groups, $K$, of size $n/k$, and for any group $S \in K$ and any agent $i \in S$ it must hold that $V_i(A_{S}) \geq \frac{k}{n}$. This condition is quite similar to $\newnotionlinearabbrev\text{-}k$; however, it is considerably less restrictive, as it places no constraints on subsets of size less than $k$, or different groupings of the agents ($\newnotionlinearabbrev\text{-}k$ considers \emph{all} possible groupings of size at most $k$).~\cite{segal2019fair} prove that unanimous-FS allocations require infinite queries; yet little is known about how close one can get with a bounded number of queries. Our algorithms provide allocations as close to $\newnotionlinearabbrev\text{-}k$ as possible, with a bounded number of queries.


Berliant et al.~\cite{berliant1992fair}, and later Husseinov~\cite{husseinov2011theory}, study a notion of group fairness, \emph{group envy-freeness}, where, similar to our work, the subsets are not predefined. However, their definition only considers groups of equal size: they call an allocation $A = (A_1, \dots, A_n)$ group envy-free, if for every pair of groups of agents $C_1$ and $C_2$, with $|C_1| = |C_2|$, there is no partition $\{ B_i \}_{i \in C_1}$ of $\cup_{j \in C_2} A_j$, such that, for all $i \in C_1$, $i$ prefers $B_i$ to $A_i$, with strict preference for at least one $i \in C_1$. This notion is stronger (``harder to achieve'') than perfection; the focus of the aforementioned works is existence of such allocations, and compatibility with economic efficiency.

Further afield, numerous papers study fair allocation of indivisible items among groups of agents~\cite{conitzer2019group,feige2022allocations,caragiannis2025new,manurangsi2022almost,scarlett2023one,kyropoulou2020almost,aziz2021almost}.

\section{Preliminaries} \label{section:prelims}



We have an infinitely divisible resource, the ``cake,'' denoted by the interval $[0,1]$. A piece of cake refers to a finite set of disjoint intervals of $[0,1]$.
Our goal is to allocate the cake among a set $N$ of $n$ agents. An allocation $\alloc = (\alloc_1, \dots, \alloc_n)$ of the cake to the agents consists of $n$ pieces, where $\alloc_i$ is the piece allocated to agent $i$, such that $\alloc_i \cap \alloc_j = \emptyset$ for all pairs of agents $i,j$. An allocation $\alloc$ is \emph{complete} if $\cup_{i=1}^n \alloc_i = [0,1]$. For an allocation $\alloc$ and a subset of agents $S \subseteq N$, we will use notation $\alloc_S = \cup_{i \in S} \alloc_i$
to denote the union of all pieces allocated to the agents in $S$. We also use notation $\bar{S} = N \setminus S$ for the complement of a set $S$.

Every agent $i \in N$ has a valuation function $V_i$ that assigns a non-negative value to any subinterval of $[0,1]$. It is convenient to think of these values as being induced by a density function $v_i$. That is, for an interval $I = [a,b]$, $V_i(I) = \int_{x=a}^b v_i(x) dx$. Valuations are (i) normalized: $V_i([0,1]) = \int_{x=0}^1 v_i(x) dx = 1$, (ii) additive: for a set of disjoint intervals $I_1, \dots, I_m$, $V_i (\cup_{j=1}^m I_j) = \sum_{j=1}^m V_i(I_j)$, (iii) non-atomic: $\forall x, y \in [0,1], \lambda \in [0,1], \exists z \in [x, y]$ where $V_i([x, z]) = \lambda \cdot V_i([x,y])$.



\subsection{Robertson-Webb Model}

We study the complexity of cake-cutting algorithms in the model suggested by Robertson and Webb~\cite{robertson1998cake} and later formalized (and named) by Woeginger and Sgall~\cite{woeginger2007complexity}. This model allows for two types of queries:
\begin{enumerate}
\item $\textsc{Eval}_i(I)$: Given an agent $i$ and an interval $I \subseteq [0,1]$, this query returns $V_i(I)$.
\item $\textsc{Cut}_i(x, v)$: Given an agent $i$, a point $x \in [0,1]$, and a value $v \in [0,1]$, this query returns the smallest point $x' \in [x, 1]$ such that $V_i([x, x']) = v$.
\end{enumerate}

To the best of our knowledge, every discrete cake-cutting algorithm can be (and has been) analyzed in this model. As a simple example, the cut-and-choose algorithm can be implemented by two queries as follows. First, a $\textsc{Cut}_1(0, 1/2)$ will return the point $y$ such that $V_1([0,y]) = V_1([y,1]) = 1/2$. The algorithm can determine whether $[0,y]$ or $[y,1]$ should be allocated to agent $1$, by making a $\textsc{Eval}_2([0,y])$ and checking whether the value returned is at most $1/2$. 

\subsection{Fairness notions}

Our goal is to produce allocations that are fair. 
An allocation $\alloc$ is \textbf{proportional} if $V_i(\alloc_i) \geq 1/n$ for all $i \in N$. An allocation $\alloc$ is \textbf{envy-free} if for all $i,j \in N$, $V_i(\alloc_i) \geq V_i(\alloc_j)$. An allocation $\alloc$ is \textbf{$\epsilon$-perfect} if for all $i,j \in N$, $1/n + \epsilon \geq V_i(\alloc_j) \geq 1/n - \epsilon$; an allocation $\alloc$ is \textbf{perfect} if it is $0$-perfect. 
Finally, an allocation, $\alloc$, is \textbf{super envy-free} if every agent values her own piece at least $1/n$, and values any other agent's piece at most $1/n$, i.e., $V_i(\alloc_i) \geq \frac{1}{n} \geq V_i(\alloc_j)$, for all $i, j \in N$.

We write $\textsc{Prop}$, $\textsc{EF}$, \textsc{Super-EF}, and $\textsc{Perf}$ for the set of all complete proportional, envy-free, super envy-free, and perfect allocations, respectively. For complete allocation, every perfect allocation is super envy-free, every super envy-free allocation is envy-free, and every envy-free allocation is proportional, and there exist perfect allocations that are not super envy-free, super envy-free allocations that are not envy-free, as well as envy-free allocations that are not proportional; that is, $\textsc{Prop} \supsetneq \textsc{EF}  \supsetneq \textsc{Super-EF} \supsetneq \textsc{Perf}$.

In this paper, we define the following new notions of fairness. 



\begin{definition}[\newnotioninverse-$k$ ($\newnotioninverseabbrev\text{-}k$)]\label{dfn: IHk}
Let $k$ be an integer, such that $n \geq k \geq 1$. An allocation $\alloc$ is \newnotioninverse-$k$ ($\newnotioninverseabbrev\text{-}k$) if, for every non-empty subset of agents $S \subseteq N$ such that $|S| \leq k$, and every $i \in S$, $V_i(\alloc_{\bar{S}}) = \sum_{j \notin S} V_i(\alloc_j) \leq \frac{n-|S|}{n-|S|+1}$.
\end{definition}

It is easy to see that the definition of \newnotioninverse-$1$ coincides with the definition of proportionality. Slightly overloading notation, we write $\newnotioninverseabbrev\text{-}k$ for the set of all complete \newnotioninverse-$k$ allocations.



\begin{definition}[\newnotionlinear-$k$ ($\newnotionlinearabbrev\text{-}k$)]\label{dfn: LHk}
Let $k$ be an integer, such that $n \geq k \geq 1$. An allocation $\alloc$ is \newnotionlinear-$k$ ($\newnotionlinearabbrev\text{-}k$) if, for every non-empty subset of agents $S \subseteq N$ such that $|S| \leq k$, and every $i \in S$, $V_i(\alloc_{\bar{S}}) = \sum_{j \notin S} V_i(\alloc_j) \leq \frac{n - |S|}{n}$.
\end{definition}

Slightly overloading notation, again, we write $\newnotionlinearabbrev\text{-}k$ for the set of all complete \newnotionlinear-$k$ allocations.
Notice that every \newnotionlinear-$k$ allocation is \newnotioninverse-$k$, since $\frac{n - |S|}{n} \leq \frac{n-|S|}{n-|S|+1}$ for all $|S| \geq 1$.

\section{Relations Between Fairness Notions}\label{sec:relations}

In this section, we show the relation between our new fairness notions, and existing fairness notions: proportionality, envy-freeness, and perfection.

\subsection{\newnotioninverseabbrev-k Relations}

\begin{theorem}\label{theorem:proptoinversektoef}
$\textsc{Prop}
= \newnotioninverseabbrev\text{-}1
\supsetneq \newnotioninverseabbrev\text{-}2
\supsetneq \dots
\supsetneq \newnotioninverseabbrev\text{-}(n-1)
= \newnotioninverseabbrev\text{-}n
\supsetneq \textsc{EF}$.
\end{theorem}

\begin{proof}[Proof of~\Cref{theorem:proptoinversektoef}]
The first equality, $\textsc{Prop}
= \newnotioninverseabbrev\text{-}1$ is immediate from the definition of $\newnotioninverseabbrev\text{-}1$: for $|S| = 1$, $1 - V_i(A_i) = V_i(A_{\bar{S}}) \leq \frac{n-1}{n} = 1 - \frac{1}{n}$. It is equally easy to see that $\newnotioninverseabbrev\text{-}k \supseteq \newnotioninverseabbrev\text{-}(k+1)$, for $1 \leq k \leq n-2$, since the conditions required to satisfy the latter are a subset of the conditions required to satisfy the former notion;~\Cref{claim:inverseknotimplyk1} gives an allocation $\alloc$ such that $\alloc \in \newnotioninverseabbrev\text{-}k$, but $\alloc \notin \newnotioninverseabbrev\text{-}(k+1)$, proving that  $\newnotioninverseabbrev\text{-}k \supsetneq \newnotioninverseabbrev\text{-}(k+1)$. 

\begin{claim} \label{claim:inverseknotimplyk1}
$\newnotioninverseabbrev\text{-}k$ does not imply $\newnotioninverseabbrev\text{-}(k+1)$ for $1 \leq k \leq n-1$.
\end{claim}
\begin{proof}
Let $\alloc \in \newnotioninverseabbrev\text{-}k$
be an allocation such that, for all $i \in N$, $V_i(\alloc_i) = \frac{1}{n-k+1}$, and there exists an agent $j_i \in N \setminus \{ i \}$, such that $V_i(\alloc_j) = 1 - \frac{1}{n-k+1}$ (all other $V_i(\alloc_w)$ are 0). It is straightforward to construct such an allocation. $\alloc \in \newnotioninverseabbrev\text{-}k$, since $V_i(A_{\bar{S}}) \leq 1 - \frac{1}{n-k+1} \leq \frac{n - |S|}{n-|S|+1}$ for all $S$ such that $|S| \leq k$ and $i \in S$. However, by picking $S$ such that $i \in S$, $j_i \notin S$, and $|S| = k+1$, we have that $V_i(A_{\bar{S}}) = 1 - \frac{1}{n-k+1} = \frac{n-k}{n-k+1} > \frac{n - |S|}{n - |S| + 1}$, i.e., $\alloc \notin \newnotioninverseabbrev\text{-}(k+1)$.
\end{proof}

The last equality, $\newnotioninverseabbrev\text{-}(n-1)
= \newnotioninverseabbrev\text{-}n$, again immediately follows from the definitions. Finally,~\Cref{claim:efimpliesinversen} proves that $\newnotioninverseabbrev\text{-}n
\supsetneq \textsc{EF}$.

\begin{claim} \label{claim:efimpliesinversen}
$\textsc{EF} \subsetneq \newnotioninverseabbrev\text{-}n$.
\end{claim}
\begin{proof}
To see that $\textsc{EF} \subseteq \newnotioninverseabbrev\text{-}n$, consider an allocation $\alloc \in \textsc{EF}$; we will show that $\alloc \in \newnotioninverseabbrev\text{-}n$. 
Let $S \subseteq N$ be an arbitrary subset of agents of size $|S| \leq n$.
By the definition of envy-freeness, we have that, for $i,j \in N$, $V_i(\alloc_j) \leq V_i(\alloc_i)$. Consider $i \in S$; adding up for all $j \notin S$ we have that 
\begin{align*}
    V_i(\alloc_{\bar{S}}) &\leq (n-|S|) V_i (\alloc_i) \\
    &\leq (n-|S|) V_i(\alloc_{S}) \tag{$V_i (\alloc_i) \leq V_i(\alloc_{S})$} \\
    &= (n-|S|)(1 - V_i(\alloc_{\bar{S}})).
\end{align*}
Re-arranging we get the desired inequality: $V_i(A_{\bar{S}}) \leq \frac{n-|S|}{n-|S|+1}$.

To see that $\newnotioninverseabbrev\text{-}n$ does not imply $\textsc{EF}$, consider an allocation $\alloc \in \newnotioninverseabbrev\text{-}n$ such that, for all agents $i \in N$ $V_i(\alloc_i) = \frac{1}{3}$; furthermore, for every agent $i \in N$, there exist distinct agents $j_i$ and $\ell_i$ such that $V_i(\alloc_{j_i}) = \frac{1}{2}$ and $V_i(\alloc_{\ell_i}) = \frac{1}{6}$ (and $V_i(\alloc_w) = 0$ for all other agents $w$). It is easy to see that such an allocation can be constructed. $\alloc \in \newnotioninverseabbrev\text{-}n$, since: (i) for all $S \subseteq N$, $|S| \leq n-2$, and all $i \in S$, $V_i(\alloc_{\bar{S}}) \leq 1 - V_i(\alloc_i) = \frac{2}{3} \leq \frac{n - |S|}{n - |S| + 1}$, (ii) for all $S \subseteq N$, $|S| = n-1$, and all $i \in S$, $V_i(\alloc_{\bar{S}}) \leq 1 - (V_i(\alloc_i) + \min\{ V_i(\alloc_{j_i)}, V_i(\alloc_{\ell_i)} \}) \leq \frac{1}{2} \leq \frac{n - |S|}{n - |S| + 1}$, and (iii) $V_i(\alloc_{\bar{S}}) = 0$ for $S = N$. However, $\alloc \notin \textsc{EF}$, since every agent $i \in N$ envies agent $j_i$.
\end{proof}

This concludes the proof of~\Cref{theorem:proptoinversektoef}.
\end{proof}

\subsection{\newnotionlinearabbrev-k Relations}

\begin{theorem}\label{thm: CLB to super-EF}
$\textsc{Prop}
= \textsc{\newnotionlinearabbrev-}1
\supsetneq \textsc{\newnotionlinearabbrev-}2
\supsetneq \dots
\supsetneq \textsc{\newnotionlinearabbrev-}(n-1)
= \textsc{\newnotionlinearabbrev-}n
= \textsc{Super-EF}$.
\end{theorem}

\begin{proof}[Proof of~\Cref{thm: CLB to super-EF}]
By the definition of \newnotionlinearabbrev-$k$, we have that $\textsc{Prop}
= \textsc{\newnotionlinearabbrev-}1$, $\textsc{\newnotionlinearabbrev-}k \supseteq \textsc{\newnotionlinearabbrev-}(k+1)$ for $1 \leq k \leq n-1$, and \newnotionlinearabbrev-$(n-1)$ is equivalent to \newnotionlinearabbrev-$n$. 
The theorem follows from~\Cref{claim:linearknotimplieslineark+1} and ~\Cref{claim:linearnequalssuperef}.

\begin{claim}\label{claim:linearknotimplieslineark+1}
\newnotionlinearabbrev-$k$ does not imply \newnotionlinearabbrev-$(k+1)$, for any $1 \leq k < n - 1$. 
\end{claim}
\begin{proof}
Consider an allocation $\alloc \in \textsc{\newnotionlinearabbrev-}k$ such that, for all $i \in N$: (i) $V_i(\alloc_i) = \frac{k}{n}$, (ii) $V_i(\alloc_{j}) = 0$, for all $j \in Z_i$, for some $Z_i \subset N \setminus \{ i \}$ such that $|Z_i| = k$, and (iii) $V_i(\alloc_{j}) = \left(1+\frac{1}{n-k-1}\right) \cdot \frac{1}{n}$, for all $j \in N \setminus ( Z_i \cup \{ i \} )$. $S_i = Z_i \cup \{ i \}$; $|S_i| = k + 1$. Then, $V_i(\alloc_{S_i}) = V_i(\alloc_i) = \frac{k}{n}$, and therefore $V_i(\alloc_{\bar{S_i}}) = \frac{n-k}{n} < \frac{n - |S_i|}{n}$, i.e. $\alloc \notin \textsc{\newnotionlinearabbrev-}(k+1)$.
\end{proof}

\begin{claim}\label{claim:linearnequalssuperef}
$\textsc{\newnotionlinearabbrev-}n = \textsc{Super-EF}$. 
\end{claim}
\begin{proof}
Consider an allocation $\alloc \in \textsc{\newnotionlinearabbrev-}n$. For all $S \subseteq N$ such that $|S| = n-1$, and all $i \in S$, $V_i(\alloc_{\bar{S}}) \leq \frac{1}{n}$; however, $\bar{S}$ has a single (arbitrary) agent. That is, $V_i(\alloc_{\bar{S}}) = V_i(\alloc_j) \leq \frac{1}{n}$, for all $j \in N \setminus \{ i \}$. And, since $\textsc{\newnotionlinearabbrev-}n \subseteq \textsc{Prop}$, we also have $V_i(\alloc_i) \geq \frac{1}{n}$, therefore, $\alloc \in \textsc{Super-EF}$.

Consider an allocation $\alloc \in \textsc{Super-EF}$. We have that, for all $i,j \in N$, $j \neq i$, $\frac{1}{n} \geq V_i(\alloc_j)$. Adding up this for all $j \notin S$, for some arbitrary set $S \subseteq N$, $|S| \leq n$, such that $i \in S$, we have $V_i(\alloc_{\bar{S}}) \leq \frac{n - |S|}{n}$, i.e., $\alloc \in \textsc{\newnotionlinearabbrev-}n$.
\end{proof}

This concludes the proof of~\Cref{thm: CLB to super-EF}
\end{proof}

\begin{theorem}\label{thm: ef and CLB}
$\textsc{EF}
\supsetneq \textsc{\newnotionlinearabbrev-}\lceil\frac{n}{2}\rceil$, but $\textsc{EF}
\not\supseteq \textsc{\newnotionlinearabbrev-}(\lceil\frac{n}{2}\rceil-1)$ and $\textsc{EF} \not\subseteq \textsc{\newnotionlinearabbrev-}2$.
\end{theorem}
\begin{proof}[Proof of~\Cref{thm: ef and CLB}]
Consider an allocation $\alloc \in \textsc{\newnotionlinearabbrev-}\lceil\frac{n}{2}\rceil$. 
Let $i,j \in N$ be two arbitrary agents. Let $S \subseteq N \setminus \{ j \}$, such that (i) $i \in S$, and (ii), $|S| = \lceil\frac{n}{2}\rceil$. We have that $V_i(\alloc_S) = 1 - V_i(\alloc_{\bar{S}}) \geq 1 - \frac{n-|S|}{n} = \frac{1}{2} + \frac{n \mod 2}{2n}$. Let $S'$ be such that $\bar{S'} = (S \setminus \{ i \} ) \cup \{ j \}$ (i.e., $S' = \{ i \} \cup ( N \setminus ( \{ j \} \cup S ) ) )$. We have that $|S'| = n - \lceil\frac{n}{2}\rceil = \lfloor\frac{n}{2}\rfloor$, therefore, since $\alloc \in \textsc{\newnotionlinearabbrev-}\lceil\frac{n}{2}\rceil$, and $i \in S'$, $V_i(\bar{S'}) \leq \frac{n - |S'|}{n} = \frac{1}{2} + \frac{n \mod 2}{2n}$. Therefore, $V_i(\alloc_S) \geq V_i(\bar{S'})$. However, $V_i(\alloc_S) = V_i(\alloc_i) + \sum_{z \in S \setminus \{ i \}} V_i(\alloc_z)$, and $V_i(\bar{S'}) = V_i(\alloc_j) + \sum_{z \in S \setminus \{ i \}} V_i(\alloc_z)$. Therefore, we have that $V_i(\alloc_i) \geq V_i(\alloc_j)$, i.e., $i$ does not envy $j$.

Next, notice that \newnotionlinearabbrev-$(\lceil\frac{n}{2}\rceil-1)$ (and by extension, \newnotionlinearabbrev-$k$ for $k < \lceil\frac{n}{2}\rceil-1$) does not imply \textsc{EF}. Consider an allocation $\alloc \in \textsc{\newnotionlinearabbrev-}(\lceil\frac{n}{2}\rceil-1)$ such that, for all $i \in N$, $V_i(\alloc_i) = (\lceil\frac{n}{2}\rceil-1) \cdot \frac{1}{n}$, and there exists an agent $j_i$ such that $V_i(\alloc_{j_i}) = 1 - (\lceil\frac{n}{2}\rceil-1) \cdot \frac{1}{n}$, and $V_i(\alloc_z) = 0$, for all agents $z \neq i,j_i$. Since $V_i(\alloc_i) < V_i(\alloc_{j_i})$, envy-freeness is violated.

Finally, \textsc{EF} does not imply \newnotionlinearabbrev-$2$ (and by extension, \newnotionlinearabbrev-$k$ for $k > 2$).
Consider an allocation $\alloc \in \textsc{EF}$ such that, for all $i \in N$ there exists an agent $j_i$, such that $V_i(\alloc_{j_i}) = 0$, and $V_i(\alloc_z) = \frac{1}{n-1}$, for all 
$z \neq i, j_i$. 
For $S = \{ i, j_i \}$ and $n > 2$, 
$V_i(A_{\bar{S}}) = \frac{n-2}{n-1}  > \frac{n-|S|}{n} = \frac{2}{n}$.
\end{proof}

\section{An Algorithm for $\newnotioninverseabbrev\text{-}n$}\label{sec: algorithms}

In this section, we prove our main upper bound: $\newnotioninverseabbrev\text{-}n$ allocations can be computed using  $O(n^4)$ queries in the Robertson-Webb model.

\begin{algorithm}[t]
\SetAlgoLined
\LinesNumbered
    Let $V^+_{n+1}, \dots, V^+_{4n/3}$ be $n/3$ arbitrary valuation functions.\\
    $B \gets \frac{1}{4n}\textsc{-Perfect}(V_1, \dots, V_n, V^+_{n+1}, \dots, V^+_{4n/3})$. \Comment{Find an $\epsilon$\textsc{-Perfect} allocation with the extra agents} \\
    $R \gets \cup_{j=n+1}^{4n/3} B_j$. \Comment{The pieces of the additional agents becomes the residue} \\
    $S \gets \{ B_1, \dots, B_n \}$, $M \gets \{ 1, \dots, n \}$. \Comment{Initialize the set of active pieces and active agents.} \\
    \While{$|M| > 1$}{
        Ask every agent $i \in M$ to make a mark on $R$ such that the piece to the left of the mark has value $1/n - \max_{B \in S} V_i(B)$.\\
        Let $i^* \in M$ be the agent with the left-most mark on $R$ (breaking ties arbitrarily). \\
        Let $R_{i^*}$ be the part of $R$ to the left of $i^*$'s mark. \\
        $\alloc_{i^*} \gets  R_{i^*} \cup \arg\max_{B \in S} V_{i^*}(B)$. \Comment{Allocate to agent $i^*$}\\
        $R \gets R \setminus R_{i^*}$. \Comment{Update the residue} \\
        $S \gets S \setminus \arg\max_{B \in S} V_{i^*}(B)$, $M \gets M \setminus \{i^* \}$. \Comment{Remove $i^*$'s piece from $S$ and $i^*$ from $M$} \\
    }
    For the remaining agent $i \in M$, $\alloc_{i} \gets R \cup S$. \\
    \Return $A$
\caption{An algorithm for finding \newnotioninverseabbrev-$n$ allocations}\label{algo:inversen}
\end{algorithm}

\begin{theorem}\label{thm: main upper bound}
\Cref{algo:inversen} computes a \textsc{\newnotioninverseabbrev}-$n$ allocation using $O(n^4)$ $\textsc{Cut}$ and $\textsc{Eval}$ queries.
\end{theorem}

\begin{proof}
We first prove that every complete and proportional allocation $\alloc$ such that  $V_i(\alloc_j) \geq \frac{1}{2n}$, for all agents $i,j$, satisfies $\newnotioninverseabbrev\text{-}n$.  

Towards proving this statement, consider such an allocation $\alloc$, and an arbitrary subset of agents $S \subseteq N$ and $i \in S$. $V_i(\alloc_{\bar{S}}) = 1 - V_i(\alloc_S) \leq  1 - \left( \frac{1}{n} + \frac{|S| - 1}{2n} \right) = \frac{2n-|S|-1}{2n}$, which is at most $\frac{n - |S|}{n - |S| + 1}$ as long as $|S| \leq n-1$. For $|S|=n$, $V_i(\alloc_{\bar{S}}) = 0 \leq \frac{n - |S|}{n - |S| + 1}$. Therefore, $\alloc \in \newnotioninverseabbrev\text{-}n$.

It remains to prove that~\Cref{algo:inversen} finds an allocation with these properties (completeness, proportionality, and $V_i(\alloc_j) \geq \frac{1}{2n}$) using $O(n^4)$ $\textsc{Cut}$ and $\textsc{Eval}$ queries.

\Cref{algo:inversen} first computes an $\frac{1}{4n}$-perfect allocation $B$ for $4n/3$ agents, where the $n/3$ extra agents, agents $n+1, \dots, 4n/3$, have arbitrary valuation functions. Let $B_1, \dots, B_{n}$ be the first $n$ pieces of this allocations; pieces $B_{n+1}, \dots, B_{4n/3}$ are combined into the residue $R$, i.e., $R = \cup_{j=n+1}^{4n/3} B_j$.
Importantly, ``real'' agents are \emph{not} allocated any of the $B_j$ pieces (yet).

\Cref{algo:inversen} then proceeds \`a la Dubins-Spanier~\cite{dubins1961cut}/Last Diminisher~\cite{steinhaus1948problem}. Let $S$ be the subset of the $B_j$ pieces still available.
The algorithm asks every agent $i$ to make a mark on $R$, such that her value to the left of the mark is equal to $1/n - \max_{B \in S} V_i(B)$. The agent $i^*$ with the left-most mark (breaking ties arbitrarily) gets the two corresponding pieces (one from cutting the residue and her favorite piece from $S$), and exits the process; we denote her final allocation by $\alloc_{i^*} = \alloc'_{i^*} \cup R_{i^*}$, where $\alloc'_{i^*}$ is her favorite piece in $S$, and (slightly overloading notation) $R_{i^*}$ is the part of the residue allocated to her. The process is repeated until a single agent $j$ is left, who is allocated the entire remaining residue, and the unique piece left in $S$.

\paragraph{Query complexity.}
Using the algorithm of Br{\^a}nzei and Miltersen~\cite{branzei2015dictatorship} for finding $\epsilon$-perfect allocation, computing $(B,R)$ costs $O( (4n/3)^3 / (1/4n) ) = O(n^4)$ queries. This protocol produces at most $O(n^2)$ intervals, and therefore, at most $O(n^3)$ \textsc{Eval} queries are necessary for every agent to compute $V_i(B_j)$. The ``\newalgosecondstage'' steps (the while loop in~\Cref{algo:inversen}) cost another $O(n^2)$ queries. Therefore, the overall complexity is $O(n^4)$.

\paragraph{Correctness.}
The final allocation is complete, by construction. First, we prove that the final allocation is proportional, i.e., $V_i(\alloc_i) \geq 1/n$. For the first $n-1$ agents to get a piece of the residue, we have, by definition, $V_i(\alloc_i) = V_i(\alloc'_i) + V_i(R_i) = \max_{B \in S} V_i(B) + (1/n - \max_{B \in S} V_i(B)) = 1/n$. For the agent $i$ that gets the last piece of the residue, every time a different agent leaves the protocol she takes with her two pieces of combined value at most $1/n$ (from the perspective of agent $i$). Therefore, the combined value of all pieces allocated is at most $(n-1)\frac{1}{n}$. Therefore, the two remaining pieces (the last $B_j$ pieces and remaining residue) have value at least $1/n$.

Regarding the second condition, $V_i(\alloc_j) \geq \frac{1}{2n}$ for all agents $i,j$, we have the following. By the fact that $B$ is $\frac{1}{4n}$-perfect for the instance with $4n/3$ agents: $V_i(B_{\ell}) \geq \frac{1}{4n/3} - \frac{1}{4n} = \frac{1}{2n}$. Observing that $V_i(\alloc_j) = V_i(\alloc'_j) + V_i(R_j) \geq V_i(\alloc'_j)$, and $\alloc'_j$ is one the $B_{\ell}$ pieces, completes the proof.
\end{proof}

\section{Lower Bounds}\label{sec: lower bounds}

In this section, we prove our lower bounds for \textsc{\newnotioninverseabbrev}-$2$ (\Cref{theorem:CHB2 lower bound}) and \textsc{\newnotionlinearabbrev}-$2$ (\Cref{theorem:LHkinfinite}).

\begin{theorem} \label{theorem:CHB2 lower bound}
Computing \textsc{\newnotioninverseabbrev}-$2$ allocations requires $\Omega(n^2)$ queries.
\end{theorem}

\begin{proof}
For this proof, we use many definitions and lemmas from the work of Procaccia~\cite{procaccia2009thou}, who proves the $\Omega(n^2)$ lower bound for finding envy-free allocations.

Consider an arbitrary algorithm.
First, to analyze the information available to the algorithm at each step, we define, for every agent $i \in N$ and every step $t$, a set of disjoint intervals $\Pi_i^t$, that is a partition of $[0,1]$. We say that interval $I \in \Pi_i^t$ is active with respect to agent $i$ at step $t$. 

$\Pi_i^t$s are defined recursively. $\Pi^0_i = \{ [0,1] \}$, since the only information available to the algorithm after $0$ steps/queries is that $V_i([0,1]) = 1$. Assuming that at step $t$ we have $\Pi_i^t$, if at step $t+1$ the algorithm does not make a query for agent $i$ (i.e., the query at step $t+1$ is $\textsc{Eval}_j$ or $\textsc{Cut}_j$ for some $j \neq i$), then $\Pi_i^{t+1} = \Pi_i^t$. Otherwise, $\Pi_i^{t+1}$ gets updated accordingly. For example, if the query is $\textsc{Eval}_i(x_1,x_2)$, where $x_1 \in I_1$ and $x_2 \in I_2$, for some intervals $I_1, I_2 \in \Pi_i^t$, then, informally,
\[
\Pi_i^{t+1} = (\Pi_i^t \setminus \{ I_1, I_2 \}) \cup \{ [left(I_1),x_1], [x_1, right(I_1)], [left(I_2),x_2], [x_2, right(I_2)] \},
\]
where for an interval $I = [a,b]$, $left(I) = a$ and $right(I) = b$. Intuitively, the algorithm at step $t$ ``knew'' $i$'s value for $I_1$ and $I_2$, and after the $\textsc{Eval}_i(x_1,x_2)$, it can infer (at most) the value of agent $i$ for four additional intervals: $[left(I_1),x_1]$, $[x_1, right(I_1)]$, $[left(I_2),x_2]$, and $[x_2, right(I_2)]$ (noting, that the first two imply the value for $I_1$ and the second two imply the value for $I_2$). Procaccia~\cite{procaccia2009thou} proves two crucial lemmas:

\begin{lemma}[\cite{procaccia2009thou}; Lemma 3.2] \label{lemma:pigrows2}
For all $i \in N$ and stage $t$, $|\Pi_i^{t+1}| - |\Pi_i^{t}| \leq 2$.
\end{lemma}

\begin{lemma}[\cite{procaccia2009thou}; Lemma 3.3]
For all $i \in N$ and stage $t$, $\Pi_i^{t}$ has the following properties:
\begin{enumerate}
    \item For every $I \in \Pi_i^{t}$, $V_i(I)$ is known to the algorithm at stage $t$.
    \item For every $I \in \Pi_i^{t}$, $I' \subsetneq I$, and $0 \leq \lambda \leq 1$, it might be the case (based on the information available to the algorithm at stage $t$) that $V_i(I') = \lambda V_i(I)$.
\end{enumerate}
\end{lemma}

Consider an adversary that, for every agent $i \in N$, responds as if the valuation of the agent was uniform over the interval $[0,1]$ (i.e., responds $\textsc{Eval}_i([x_1, x_2]) = x_2 - x_1$, and $\textsc{Cut}_i(x,v) = x + v$). Let $T$ be the number of queries our algorithm asks before it terminates and outputs allocation $\alloc = (\alloc_1, \dots, \alloc_n) \in \textsc{\newnotioninverseabbrev}\text{-}2$. 

First, we claim that for all $i \in N$, there exists an active interval $I_i \subseteq \alloc_i$, such that (i) $I_i \in \Pi^{T}_i$, and (ii) $V_i(I_i) \geq \frac{1}{n}$. If this is not the case, one can define $V_i$ (consistently with $\Pi^{T}_i$) such that proportionality is violated. Concretely, (1) if $I \notin \Pi^{T}_i$ for all $I \subseteq \alloc_i$ then we can concentrate the value of all $I$ such that $I \cap \alloc_i \neq \emptyset$ to the sub-intervals outside of $\alloc_i$ (and therefore, $V_i(\alloc_i) = 0$), and (2) if $V_i(I) < \frac{1}{n}$ for all $I \subseteq \alloc_i$, $I \in \Pi^{T}_i$, then we can pick $V_i$ such that $V_i(\alloc_i)= \max_{I \in \Pi^{T}_i} V_i(I) < 1/n$.

Since all queries until time $T$ have been answered as if the valuations were uniform, it must be that $1/n \leq V_i(I_i) = |I_i|$ for all $i \in N$. However, if for all $i \in N$ we have: (i) $I_i \subseteq \alloc_i$, (ii) $|I_i| \geq 1/n$, (iii) $\alloc_i \cap \alloc_j = \emptyset$ for all $j \neq i$, and (iv) $\cup_{i=1}^n \alloc_i = [0,1]$, then it must be that $|I_i| = |\alloc_i| = 1/n$, as well as $V_i(\alloc_i) = 1/n$, for all $i \in N$.
The \textsc{\newnotioninverseabbrev}-$2$ property then implies that $V_i(\alloc_{N \setminus \{ i , j\}}) = 1 - V_i(\alloc_i) - V_i(\alloc_j) \leq 1 - \frac{1}{n-1}$, or $V_i(\alloc_j) \geq \frac{1}{n-1} - \frac{1}{n} = \frac{1}{n(n-1)}$ for all $i, j \in N$.

To meet this condition, it must be that for every agent $i$ and piece $\alloc_j$, there exists an active interval $I \in \Pi^{T}_i$ such that (i) $I \subseteq \alloc_j$, and (ii) $V_i(I) \neq 0$. Similarly to our earlier argument, if this is not the case, one can define $V_i$ (consistently with $\Pi^{T}_i$) such that $V_i(\alloc_j) = 0$ ($< \frac{1}{n(n-1)}$). Concretely, we can pick $V_i$ such that $V_i(I \cap A_j) = 0$ and $V_i(I \setminus A_j) = V_i(I)$ for all $I \in \Pi^T_i$ and $I \cap A_j \neq \emptyset$.

Since allocations are pair-wise disjoint, we must then have at least $n$ active intervals per agent by time $T$, i.e. $|\Pi^T_i| \geq n$ for every agent $i$. Since $|\Pi_i^0| = 1$,~\Cref{lemma:pigrows2} implies that $|\Pi_i^T| \leq 2T + 1$, and therefore  $2T+1 \geq n$, i.e. the algorithm makes at least $\frac{n-1}{2}$ queries to every agent $i \in N$. Overall, the algorithm makes at least $n \cdot \frac{n-1}{2} \in \Omega(n^2)$ queries overall.
\end{proof}

\begin{theorem} \label{theorem:LHkinfinite}
Computing \textsc{\newnotionlinearabbrev}-$2$ allocations requires an infinite number of queries, for all $n \geq 3$.
\end{theorem}
\begin{proof}
We prove the statement for all odd $n$, $n \geq 3$; the proof can be easily adjusted to even $n$.

Consider an instance where $\lfloor\frac{n}{2}\rfloor$ agents --- agents $1$ through $\lfloor\frac{n}{2}\rfloor$ --- have the same valuation function $V_1$, $\lfloor\frac{n}{2}\rfloor$ agents --- agents $\lfloor\frac{n}{2}\rfloor + 1$ through $n-1$ --- have the same valuation function $V_2$, and the last agent, agent $n$, has valuation function $V_n(x) = \frac{V_1(x) + V_2(x)}{2}$ for all $x \subseteq [0,1]$. 

Let $\alloc$ be a $\newnotionlinearabbrev\text{-}2$ allocation. For every agent $i = 1, \dots, \lfloor\frac{n}{2}\rfloor$, let $j_i = i + \lfloor\frac{n}{2}\rfloor$. Since agent $i$ has valuation $V_1$, we have that $V_1([0,1] \setminus (\alloc_i \cup \alloc_{j_i})) \leq 1 - 2/n$, or simply, $V_1(\alloc_i \cup \alloc_{j_i}) \geq 2/n$, from the definition of $\newnotionlinearabbrev\text{-}2$. It also holds that $V_2(\alloc_i \cup \alloc_{j_i}) \geq 2/n$. We have $\cup_{i=1}^{\lfloor\frac{n}{2}\rfloor} \left( \alloc_i \cup \alloc_{j_i} \right) = [0,1] \setminus \alloc_n$, and $V_1([0,1]) = V_2([0,1]) = 1$, therefore, $V_1(\alloc_n) = 1 - \sum_{i=1}^{\lfloor\frac{n}{2}\rfloor} V_1(\alloc_i \cup \alloc_{j_i}) \leq 1 - \lfloor\frac{n}{2}\rfloor \frac{2}{n} = 1 - \frac{n-1}{2} \frac{2}{n} = \frac{1}{n}$. Similarly, $V_2(\alloc_n) \leq \frac{1}{n}$.
Since $\newnotionlinearabbrev\text{-}2$ implies proportionality, $V_n(\alloc_n) = \frac{V_1(\alloc_n) + V_2(\alloc_n)}{2} \geq \frac{1}{n}$, i.e., $V_1(\alloc_n) + V_2(\alloc_n) \geq \frac{2}{n}$. Therefore, it must be that $V_1(\alloc_n) = V_2(\alloc_n) = \frac{1}{n}$.

Since $V_1(\alloc_i \cup \alloc_{j_i}) \geq \frac{2}{n}$, for all $i = 1, \dots, \lfloor\frac{n}{2}\rfloor$, and $\lfloor\frac{n}{2}\rfloor \frac{2}{n} = \frac{n-1}{2} \frac{2}{n} = 1 - \frac{1}{n}$ (which is exactly the value of $[0,1] \setminus \alloc_n$), it must be that $V_1(\alloc_i \cup \alloc_{j_i}) = \frac{2}{n}$. Similarly, it must be that $V_2(\alloc_i \cup \alloc_{j_i}) = \frac{2}{n}$.

Therefore, we overall have that, there are $\lfloor\frac{n}{2}\rfloor + 1$ pieces, $\alloc_n$ and the $(\alloc_i \cup \alloc_{j_i})$s, such that, $V_1$ and $V_2$ have the same value of $\frac{1}{n}$ for the first piece, and the same value of $\frac{2}{n}$ for all remaining pieces.
Given $n$ valuations functions $U_1, \dots, U_n$ and weights $w_1, \dots, w_k$, asking for a partition of the cake into $k$ pieces $I_1, \dots, I_k$ such that $U_i(I_j) = w_j$, is known as the \emph{exact division}, or consensus splitting, problem. This problem is known to be impossible to solve with a bounded protocol, even for the case of two valuation functions and $k=2$ pieces with equal weights~\cite{robertson1998cake}. Since finding a \textsc{\newnotionlinearabbrev}-$2$ allocation would give a solution to the exact division problem for two valuation functions and $\lfloor\frac{n}{2}\rfloor + 1$ pieces, we can conclude that there is no bounded protocol for finding a \textsc{\newnotionlinearabbrev}-$2$  allocation.

Note that for even $n$, the proof is near-identical: the only difference is that the $n$-th agent is not necessary (i.e., half the agents have valuation $V_1$ and the other half have valuation $V_2$), and using the same arguments we would conclude that $V_1(\alloc_i \cup \alloc_{j_i}) = V_2(\alloc_i \cup \alloc_{j_i}) = \frac{2}{n}$, for all $i = 1, \dots, n/2$.
\end{proof}

\section{Algorithms for Relaxations of \textsc{\newnotionlinearabbrev}-$n$}
\label{sec: approximations}

In this section, we study approximations of \textsc{\newnotionlinearabbrev}-$k$, aiming to bypass the strong lower bounds of~\Cref{theorem:LHkinfinite}.
Since \textsc{\newnotionlinearabbrev}-$2$ is impossible, but \textsc{\newnotionlinearabbrev}-$1$ (aka, proportionality) is easy to achieve, the minimally weaker notion that we could hope to achieve is the following: (i) if $|S| = 1$, then $V_i(\alloc_{\bar{S}}) \leq \frac{n-1}{n}$, for all $i \in S$ (i.e., the allocation is proportional), (ii) if $|S|=2$, then $V_i(\alloc_{\bar{S}}) \leq \frac{n-2}{n}(1 + \delta)$, for all $i \in S$. More generally, we have

\begin{definition}[$\delta$-\textsc{\newnotionlinearabbrev}-$k$]
Let $k$ be an integer, such that $n \geq k \geq 1$. An allocation $\alloc$ is $\delta$-\newnotionlinear-$k$ ($\newnotionlinearabbrev\text{-}k$) if it is proportional, and, for every non-empty subset of agents $S \subseteq N$ such that $2 \leq |S| \leq k$, and every $i \in S$, $V_i(\alloc_{\bar{S}}) = \sum_{j \notin S} V_i(\alloc_j) \leq \frac{n - |S|}{n} (1 + \delta)$.
\end{definition}

We prove that $\delta$-\textsc{\newnotionlinearabbrev}-$n$ allocations can be computed efficiently, by expanding on our approach for \textsc{\newnotioninverseabbrev}-$n$, i.e.,~\Cref{algo:inversen}. First, it is easy to see that a complete, $\epsilon$-perfect, and proportional allocation $\alloc$ is $(\epsilon n)$-\textsc{\newnotionlinearabbrev}-$n$: for all $S \subseteq N$ such that $2 \leq |S|$, and $i \in S$, $V_i(\alloc_{\bar{S}}) \leq (n-|S|)\left(\frac{1}{n} + \epsilon\right) = \frac{n - |S|}{n} (1 + \epsilon n)$.
We give an algorithm,~\Cref{algo:almostlinearn}, for finding a complete, $\epsilon$-perfect, and exactly proportional allocation that requires $O\left(\frac{n^5}{\epsilon} \, \frac{\ln(1/\epsilon)}{\ln(n)} \right)$ queries, which might be on independent interest. To the best of our knowledge, the closest results in the literature are: (i) an algorithm that finds proportional and $\epsilon$-equitable allocations using $O(n( \log n + \log (1 / \epsilon))$ queries, by Cechl{\'a}rov{\'a} and Pill{\'a}rov{\'a}~\cite{cechlarova2012computability}, and (ii) the $O(n^3/\epsilon)$ algorithm of Br{\^a}nzei and Miltersen~\cite{branzei2015dictatorship} for finding $\epsilon$-perfect allocations.
~\Cref{algo:almostlinearn} immediately implies that the query complexity of finding a $\delta$-\textsc{\newnotionlinearabbrev}-$n$  allocation is $O\left(\frac{n^6}{\delta} \, \frac{\ln(1/\delta)}{\ln(n)} \right)$.

\begin{theorem}\label{thm: extra upper bound}
For all $\epsilon > 0$, there exists $d \in \Theta\left( \frac{\ln(1/\epsilon)}{\ln(n)} \right)$ and $\epsilon' \in \Theta\left(\frac{\epsilon}{n^2}\right)$, such that~\Cref{algo:almostlinearn}, with parameters $d$ and $\epsilon'$, computes a complete, $\epsilon$-perfect, and proportional allocation using $O\left(\frac{n^5}{\epsilon} \, \frac{\ln(1/\epsilon)}{\ln(n)} \right)$ $\textsc{Cut}$ and $\textsc{Eval}$ queries.
\end{theorem}

\Cref{algo:almostlinearn} differs from~\Cref{algo:inversen} in its computation of $S$ and the residue $R$ prior to the final ``\newalgosecondstage'' phase (i.e. line 5 in~\Cref{algo:inversen}, line 11 in~\Cref{algo:almostlinearn} ).
In~\Cref{algo:inversen}, the goal was for each piece of $S$ to have value at least $\frac{1}{2n}$, but at most $\frac{1}{n}$ for any agent.
In~\Cref{algo:almostlinearn}, the goal is similar, but quite harder. Each piece of $S$ must have value at least $\frac{1}{n} - \tilde{\epsilon}$ but at most $\frac{1}{n}$ for any agent, for $\tilde{\epsilon} = \frac{\epsilon}{n}$; as we show, this will suffice for $\epsilon$-perfection. 

To achieve this~\Cref{algo:almostlinearn} computes $S$ and $R$ recursively. Starting with the entire cake, it computes an $(n + 1)$-piece $\epsilon'$-perfect allocation, designating a single piece of the allocation as $R$ and the remaining $n$ pieces as $S$.
Then it repeats this process, treating the residue $R$ itself as an entire cake, designating a piece of the new allocation as the new residue, and combining each of the remaining $n$ pieces with a distinct piece in $S$. 
This process is done $d$ times in total, with the goal that the expected value of each piece of $S$ (for any agent) is spaced enough between $\frac{1}{n} - \tilde{\epsilon}$ and $\frac{1}{n}$. This allows us to choose a non-zero $\epsilon'$ that keeps the potential values of each piece in $S$ for any agent within a desired range. In short, if the number of subdivisions $d$ and the margin of error for near-perfect divisions $\epsilon'$ are chosen correctly, then $\frac{1}{n} - \tilde{\epsilon}\leq V_i(B_j) \leq \frac{1}{n}$ for all $i, j \in N$.

\begin{algorithm}[ht]
\SetAlgoLined
\LinesNumbered
    \KwIn{Parameters $\epsilon'$ and $d$.}
    Let $V^+$ be an arbitrary valuation function.\\
    Let $S$ be a set of $n$ empty pieces. 
    $R \gets \{[0, 1]\}$ \Comment{Initialize the residue as the entire cake}

    \For{$t \in [1, d]$}{
        
        $B \gets \epsilon'\textsc{-Perfect}(V_1, \dots, V_n, V^+)$ with respect to, and normalized on, $R$.\\

        
        $R \gets B_{n+1}$

        \For{$i \in [1, n]$}{
        $S_i \gets S_i \cup B_i$
        \Comment{ Assign the other $n$ pieces in $B$ to a piece in $S$}
        }
    }

    $M \gets \{ 1, \dots, n \}$ \\

    \While{$|M| > 1$}{
        Ask every agent $i \in M$ to make a mark on $R$ such that the piece to the left of the mark has value $1/n - \max_{B \in S} V_i(B)$.\\
        Let $i^* \in M$ be the agent with the left-most mark on $R$ (breaking ties arbitrarily). \\
        Let $R_{i^*}$ be the part of $R$ to the left of $i^*$'s mark. \\
        $\alloc_{i^*} \gets  R_{i^*} \cup \arg\max_{B \in S} V_{i^*}(B)$. \Comment{Allocate to agent $i^*$}\\
        $R \gets R \setminus R_{i^*}$. \Comment{Update the residue} \\
        $S \gets S \setminus \arg\max_{B \in S} V_{i^*}(B)$, $M \gets M \setminus \{i^* \}$. \Comment{Remove $i^*$'s piece from $S$ and $i^*$ from $M$} \\
    }
    For the remaining agent $i \in M$, $\alloc_{i} \gets R \cup S$. \\
    \Return $A$
\caption{An algorithm for finding complete, $\epsilon$-perfect, and exactly proportional allocations}\label{algo:almostlinearn}
\end{algorithm}

\begin{proof}[Proof of~\Cref{thm: extra upper bound}]
We show that if the parameters $\epsilon'$ and $d$ are picked correctly, ~\Cref{algo:almostlinearn} outputs an allocation where $V_i(A_i) \geq \frac{1}{n}$, $V_i(A_j) \geq \frac{1}{n} - \tilde{\epsilon}$, $\forall i, j \in N$ for any $\tilde{\epsilon} > 0$.
By picking $\tilde{\epsilon} = \frac{\epsilon}{n}$, our allocations are $\epsilon$-perfect (in addition to proportional and complete). To see this, notice that (i) $V_i(A_j) \geq \frac{1}{n} - \tilde{\epsilon} = \frac{1}{n} - \frac{\epsilon}{n} \geq \frac{1}{n} - \epsilon$, and (ii) $V_i(A_j) \leq 1 - (n-1)\left( \frac{1}{n} - \tilde{\epsilon} \right) = 1 - (n-1)\left( \frac{1}{n} - \frac{\epsilon}{n} \right) = \frac{1}{n} + \epsilon - \frac{\epsilon}{n} \leq \frac{1}{n} + \epsilon$.
In the remainder of the proof, we prove that~\Cref{algo:almostlinearn} satisfies the desired property, and has runtime $O\left(\frac{n^4}{\tilde{\epsilon}} \cdot \frac{\ln(1/\tilde{\epsilon})}{\ln(n)}\right)$.

\paragraph{Picking parameters $\epsilon'$ and $d$.} 
Consider the partition, $B$, created in the $t^{th}$ iteration of line 4. Due to the normalization of $R$ at each iteration, $B$ will have the following property:
$$\left( \frac{1}{n+1} - \epsilon' \right)^t
\leq V_i(B_j)
\leq \left( \frac{1}{n+1} + \epsilon' \right)^t,
\: \forall i, j \in N$$
Let $S'$ describe the state of $S$ prior to the final ``\newalgosecondstage''~phase of the algorithm (i.e. line 11). $S'_i$ will be comprised of 1 piece from each $\epsilon'$-perfect partition, $B$, created. As discussed, each piece of $S'$ needs to have value at least $\frac{1}{n} - \tilde{\epsilon}$ but at most $\frac{1}{n}$ for any agent.
$$ V_i(S'_j) \leq \sum_{t=1}^d \left( \frac{1}{n+1} + \epsilon' \right)^t \leq \frac{1}{n} $$
$$ V_i(S'_j) \geq \sum_{t=1}^d \left( \frac{1}{n+1} - \epsilon' \right)^t \geq \frac{1}{n} - \tilde{\epsilon} $$
Notice that by polynomial expansion, we can see that:
$$ \left( \frac{1}{n+1} + \epsilon' \right)^t - \left( \frac{1}{n+1} \right)^t
\geq \left( \frac{1}{n+1} \right)^t - \left( \frac{1}{n+1} - \epsilon' \right)^t $$
Therefore, if we define $d$ such that $\sum_{t=1}^d \left( \frac{1}{n+1} \right)^t \geq \frac{1}{n} - \frac{\tilde{\epsilon}}{2}$ (the average of the required upper and lower bounds of $V_i(S'_j)$), then for increasing $\epsilon'$, the computed upper bound of $V_i(S'_j)$ will exceed $\frac{1}{n}$ before the computed lower bound dips below $\frac{1}{n} - \tilde{\epsilon}$. With this in mind, we define $d$ as described:
$$\sum_{t=1}^d \left( \frac{1}{n+1} \right)^t
= \frac{1 - \left( \frac{1}{n+1} \right)^d}{n} 
\geq \frac{1}{n} - \frac{\tilde{\epsilon}}{2}$$
$$
d 
= \lc \frac{\ln\left(\frac{\tilde{\epsilon} n}{2}\right)}{\ln\left( \frac{1}{n+1} \right)} \rc
= \lc \frac{\ln(2) + \ln(1/\tilde{\epsilon}) + \ln(1/n)}{\ln(n+1)} \rc
\in \Theta\left( \frac{\ln(1/\tilde{\epsilon})}{\ln(n)} \right).
$$
Now, we define $\epsilon'$ such that $V_i(S'_j) \leq \frac{1}{n}$, noting that if the required upper bound holds, then so will the required lower bound. For simplicity, we first loosen the upper bound:
\begin{align*}
V_i(S'_j) &\leq \sum_{t=1}^d \left( \frac{1}{n+1} + \epsilon' \right)^t\\
&= \left( \frac{1}{n+1} + \epsilon' \right) \cdot \frac{1 - \left( \frac{1}{n+1} + \epsilon' \right)^d}{1 - \left( \frac{1}{n+1} + \epsilon' \right)} \\
&\leq \left( \frac{1}{n+1} + \epsilon' \right) \cdot \frac{1 - \left( \frac{1}{n+1} \right)^{d}}{1 - \left( \frac{1}{n+1} + \epsilon' \right)}\\
&\leq \left( \frac{1}{n+1} + \epsilon' \right) \cdot \frac{1 - \frac{1}{n+1} \cdot \frac{\tilde{\epsilon} n}{2}}{1 - \left( \frac{1}{n+1} + \epsilon' \right)}.
\end{align*} 
\noindent
Enforcing that this last expression is at most $1/n$ we have $\frac{\left( \frac{1}{n+1} + \epsilon' \right)}{1 - \left( \frac{1}{n+1} + \epsilon' \right)} \leq \frac{1}{n \left( 1 - \frac{1}{n+1} \cdot \frac{\tilde{\epsilon} n}{2} \right)}$, or
\[
\epsilon' \leq \frac{1}{n \left( 1 - \frac{1}{n+1} \cdot 
 \frac{\tilde{\epsilon} n}{2} \right) + 1} - \frac{1}{n+1}
= \frac{\tilde{\epsilon} n^2}{2(n+1)^2\left( n + 1 -  \frac{\tilde{\epsilon} n^2}{2(n+1)} \right)}. 
\]
Considering that $\frac{\tilde{\epsilon} n^2}{2(n+1)} > 0$, we can tighten this constraint by decreasing the upper bound as so:
\[
\epsilon' \leq \frac{\tilde{\epsilon} n^2}{2(n+1)^3}
\in \Theta\left(\frac{\tilde{\epsilon}}{n}\right).
\]
Thus, the aforementioned conditions will hold for some $d \in \Theta\left( \frac{\ln(1/\tilde{\epsilon})}{\ln(n)} \right)$ and $\epsilon' \in \Theta\left(\frac{\tilde{\epsilon}}{n}\right)$.

\paragraph{Query Complexity.} We claim that the algorithm requires at most $O\left(\frac{n^4}{\tilde{\epsilon}} \cdot \frac{\ln(1/\tilde{\epsilon})}{\ln(n)}\right)$ to compute an allocation such that $V_i(A_i) \geq \frac{1}{n}$, and $V_i(A_j) \geq \frac{1}{n} - \tilde{\epsilon}$. The algorithm involves (i) computing $d$ $(n+1)$-piece $\epsilon'$-perfect allocations and (ii) performing a ``cut-and-match \newalgosecondstage.'' Br{\^a}nzei and Miltersen give a protocol for finding $\epsilon$-perfect allocations that uses $O\left(\frac{n^3}{\epsilon}\right)$ queries ~\cite{branzei2015dictatorship}, and it is well-known that traditional ``\newalgosecondstage'' requires $O(n^2)$ queries.
However, it is premature to conclude that this would require $O\left(\frac{n^3}{\epsilon'} \right) \cdot d + O(n^2) \in O\left(\frac{n^3}{\epsilon'} \cdot d\right)$ queries. Many of these ``queries'' will be performed over a collection of noncontiguous intervals, rather than a contiguous cake; we call these super-queries. To account for this issue, we prove that evaluations of noncontiguous intervals for each agent, the upper bound on the number of queries required of this algorithm remains $O\left(\frac{n^3}{\epsilon'} \cdot d\right)$.

Let $\mathcal{I} = \{ I_1, \dots, I_z \}$ be a collection of noncontiguous intervals. First, we show that if every agent's evaluation of each noncontiguous interval is known, each super-query requires at most $2$ actual queries. Consider the super-query $\textsc{Eval}_i([x_1, x_2])$ where $x_1 \in I$ and $x_2 \in I'$ for some intervals $I, I' \in \mathcal{I}$. If $I = I'$, $x_1$ and $x_2$ belong to the same interval, meaning only 1 actual query is required. If $I \neq I'$, then $\textsc{Eval}_i([x_1, x_2]) = \textsc{Eval}_i([x_1, I_{\text{right}}]) + \textsc{Eval}_i([I'_{\text{left}}, x_2]) + \sum_{I \in \mathcal{I}: I \text{ ``between'' } I, I'} V_i(I)$.
Now consider the super-query $\textsc{Cut}_i(v_1, x_1)$ where $x_1 \in I$ for some interval $I \in \mathcal{I}$. This query can be answered using at most 2 queries via the following process: First, perform the true query $\textsc{Eval}_i([x_1, I_{\text{right}}])$. Then, using this evaluation along with those of the intervals to the right of $I$ to determine the interval $I'$ the mark $x_2$ will be placed and the value $v_2 = v_1 - V_i(x_1, I'_{\text{left}})$ that should be to the left of $x_2$ in $I'$. Finally, only a single actual query is required to determine the cut $x_2$, $x_2 = \textsc{Cut}_i(v_2, I_{2, \text{left}})$. Thus, given each agent's evaluation of each noncontiguous interval, any super-query requires at most $2$ actual queries.  

Next, we show that if we tracked each agent's evaluation of each noncontiguous interval, then the algorithm would still require just $O\left(\frac{n^3}{\epsilon'} \cdot d\right)$ (actual) queries. Prior to computing each $\epsilon'$-perfect allocation, we must normalize each agent's valuation function over $R$. While we do not have access to valuation functions, this effect can be achieved by scaling the output of an $\textsc{Eval}$ super-query by $\frac{1}{V_i(R)}$, or scaling the input value of a $\textsc{Cut}$ super-query by $V_i(R)$. Fortunately, since $R$ is a set of intervals whose values are assumed to be known, this requires no additional queries. For the given information, we have shown that each super-query needs at most two actual queries. Thus, computing all $d$ $\epsilon'$-perfect allocations (with $(n+1)$-pieces each) requires $O\left(\frac{n^3}{\epsilon'} \cdot d\right)$ (actual) queries. As for performing our ``cut-and-match Last Diminisher'', at worst, an agent will perform $\textsc{Eval}$ super-queries on every piece in $S$ and $n$ \textsc{Cut} super-queries on $R$. Applying our prior result again, this translates to only $O(n^2)$ actual queries in total (which is dominated by $O\left(\frac{n^3}{\epsilon'} \cdot d\right)$). Thus, the aforementioned claim holds.

Finally, we show that tracking each agent's evaluation of each noncontiguous interval requires $O(n^3 \cdot d)$ queries (which is also dominated by $O\left(\frac{n^3}{\epsilon'} \cdot d\right)$). Consider that the $\epsilon$-perfect algorithm by Br{\^a}nzei and Miltersen~\cite{branzei2015dictatorship} produces allocations with $O(n^2)$ cuts (thus introducing $O(n^2)$ noncontiguous intervals). To incorporate each new cut $x$ on an existing interval $I$ (thus creating intervals $I^+_1 = [I_{\text{left}}, x]$ and $I^+_2 = [x, I_{\text{right}}]$) for an agent $i$, we perform 1 query $\textsc{Eval}_i(I^+_1)$ to determine the values of the new intervals created ($V_i(I^+_2)$ is deduced from $V_i(I) - V_i(I^+_1)$). Thus, introducing a cut (and by extension a noncontiguous interval) only requires $O(n)$ true queries to keep our noncontiguous interval evaluations up-to-date. Since our $d$ $\epsilon'$-perfect allocations introduce $O(n^2 \cdot d)$ cuts and our ``cut-and-match \newalgosecondstage'' introduces $O(n)$ cuts (one for each of the first $n-1$ pieces allocated), tracking the desired information requires $O(n^3 \cdot d) + O(n) \in O(n^3 \cdot d)$ queries in total.

Putting these pieces together, we can conclude that~\Cref{algo:almostlinearn} computes complete and proportional allocations, where $V_i(A_j) \geq \frac{1}{n} - \tilde{\epsilon}$ for all $i, j \in N$, using $O\left( \frac{n^3}{\epsilon'} \cdot d \right) = O\left(
    \frac{n^4}{\tilde{\epsilon}} \cdot \frac{\ln(1/\tilde{\epsilon})}{\ln(n)}
\right)$ queries in total (and therefore a complete, proportional and $\epsilon$-perfect allocation using $O\left(
    \frac{n^5}{\epsilon} \cdot \frac{\ln(1/\epsilon)}{\ln(n)}
\right)$ queries).



\paragraph{Correctness.} 
From our earlier discussion on query complexity and choice of parameters $d$ and $\epsilon'$, we have that the first phase of the algorithm (lines 1 to 10) end with $n$ pieces, $S_1, \dots, S_n$, such that $\frac{1}{n} -\tilde{\epsilon} \leq V_i(S_j) \leq \frac{1}{n}$, as well as a residue. Since the final allocation of every agent $j$ is a superset of some piece $S_{\ell}$, we have that $V_i(\alloc_j) \geq V_i(S_{\ell}) \geq \frac{1}{n} - \tilde{\epsilon}$. For the first $n-1$ agents to exit the protocol (i.e., the agents that are allocated during lines 12 through 19), $V_i(\alloc_i) \geq \frac{1}{n}$, by definition. For the remaining agent, every pair of pieces allocated so far (an $S_j$ piece and a piece of the residue) has value at most $\frac{1}{n}$, and there are at most $n-1$ pairs, therefore $V_i(\alloc_i) \geq \frac{1}{n}$ for this agent as well.
\end{proof}

\newpage
\section*{Acknowledgements}


The authors would like to thank Yorgos Amanatidis, Yorgos Christodoulou, John Fearnley, and Vangelis Markakis for their input in the initial development of this project. The authors would also like to thank Ariel Procaccia for the valuable discussions and suggestions.

\bibliographystyle{alpha}
\bibliography{refs}




\end{document}